\documentclass[conference]{IEEEtran}
\IEEEoverridecommandlockouts
\usepackage{cite}
\usepackage{amsmath,amssymb,amsfonts}
\usepackage{algorithm}
\usepackage{algpseudocode}
\usepackage{amsmath} % For math formatting
\usepackage{graphicx}
\usepackage{textcomp}
\usepackage{xcolor}
\def\BibTeX{{\rm B\kern-.05em{\sc i\kern-.025em b}\kern-.08em
    T\kern-.1667em\lower.7ex\hbox{E}\kern-.125emX}}
\begin{document}

\title{Multithreaded Fine-Grained Asynchronous BSP for Integer Sorting with LCI and OpenMP}

\author{\IEEEauthorblockN{Minyu Cheng}
\IEEEauthorblockA{\textit{Siebel School of Comp. \& Data Science} \\
\textit{University of Illinois Urbana-Champaign}\\
Urbana, IL, USA \\
minyuc2@illinois.edu}
\and
\IEEEauthorblockN{Jiakun Yan\IEEEauthorrefmark{1}\thanks{\IEEEauthorrefmark{1}This work was done when J. Yan was with University of Illinois Urbana-Champaign.}}
\IEEEauthorblockA{\textit{Advanced Micro Devices, Inc.}\\
Bellevue, WA, USA \\
jiakun.yan@amd.com}
\and
\IEEEauthorblockN{Marc Snir}
\IEEEauthorblockA{\textit{Siebel School of Comp. \& Data Science} \\
\textit{University of Illinois Urbana-Champaign}\\
Urbana, IL, USA \\
snir@illinois.edu}
}

\maketitle

\begin{abstract} 
    The bulk synchronous parallel (BSP) model struggles with irregular workloads due to rigid global communication. While fine-grained asynchronous BSP (FA-BSP) improves overlap, existing implementations typically rely on a limiting one-process-per-core model. This paper proposes a multithreaded FA-BSP approach combining Lightweight Communication Interface (LCI) and OpenMP to fully exploit multicore architectures. We evaluate this design using the NAS Parallel Benchmark Integer Sort (IS), retaining the original irregular Gaussian distribution to rigorously test load balancing. By replacing synchronous MPI collectives with OpenMP multithreading and LCI's fine-grained, zero-copy active messages, we enable efficient computation-communication overlap. Our evaluation demonstrates that multithreaded FA-BSP significantly outperforms traditional bulk-synchronous MPI implementations, offering a scalable solution for irregular scientific applications. 
\end{abstract}


% All figures for the paper

\newcommand{\FigOverall}{
\begin{figure}[h]
\centering
\includegraphics[width=\columnwidth]{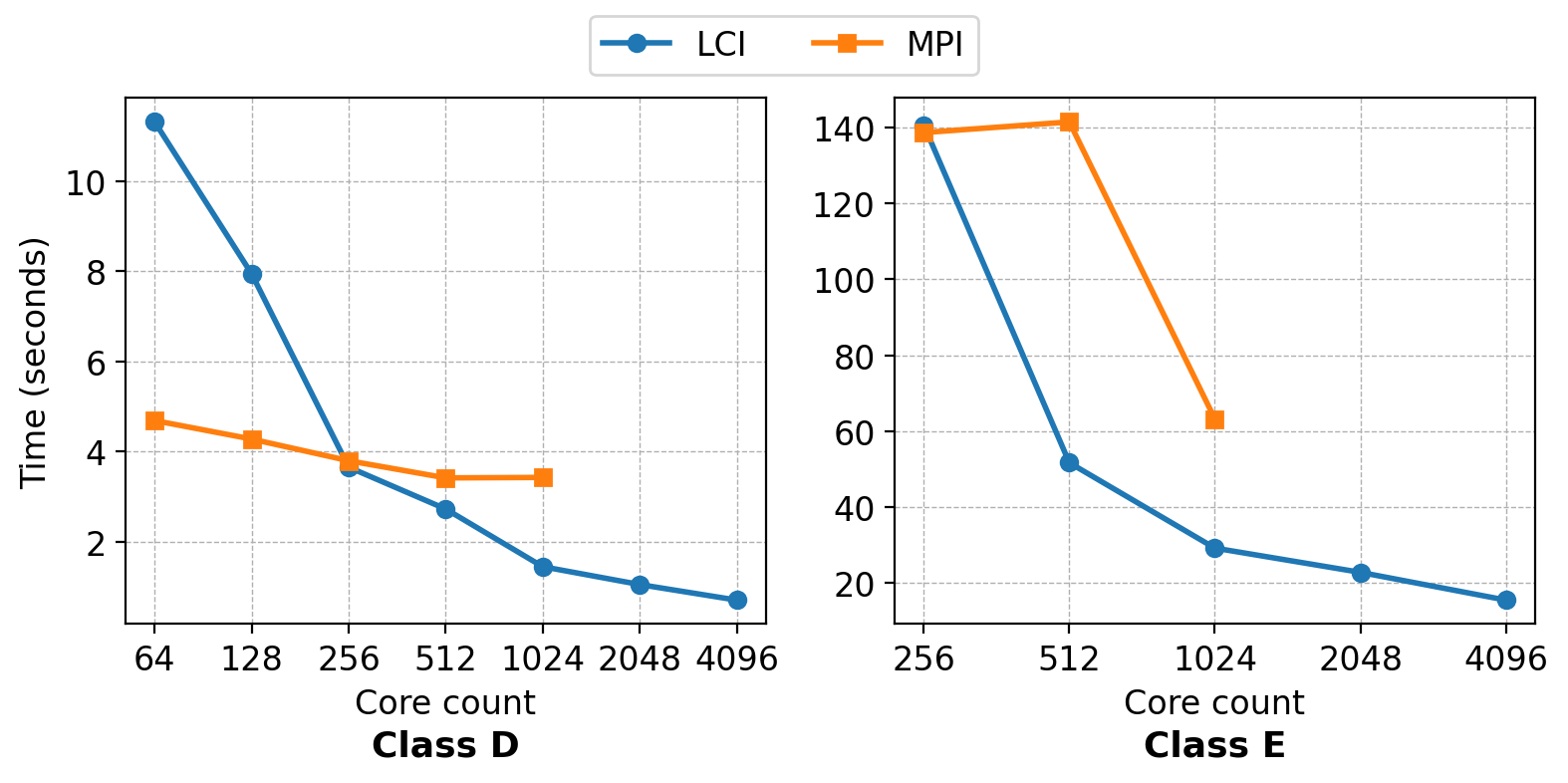}
\caption{Execution time (median over 5 runs) comparison between LCI and MPI implementations against Class D and E, across different core counts.}
\label{fig:overall}
\end{figure}
}

\newcommand{\FigDeviceCount}{
\begin{figure}[h]
\centering
\includegraphics[width=0.95\columnwidth]{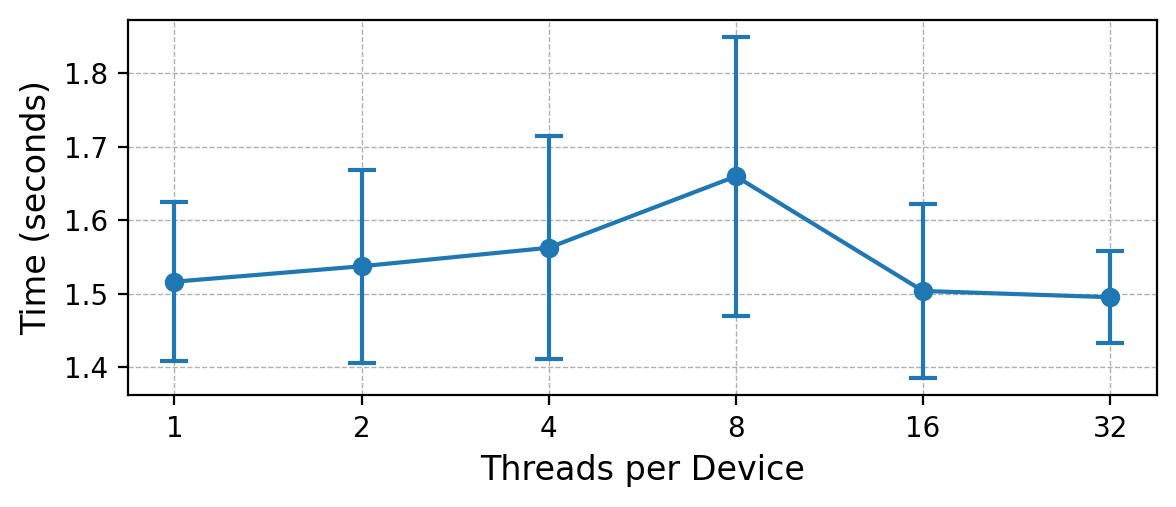}
\caption{Execution time (average over 5 runs) versus number of threads per device under Class D, with a fixed configuration of 32 processes and 32 threads per process (1024 cores total). Error bars show the standard deviation.}
\label{fig:device_count}
\end{figure}
}

\newcommand{\FigProcessWidth}{
\begin{figure*}[t]
\centering
\includegraphics[width=\textwidth]{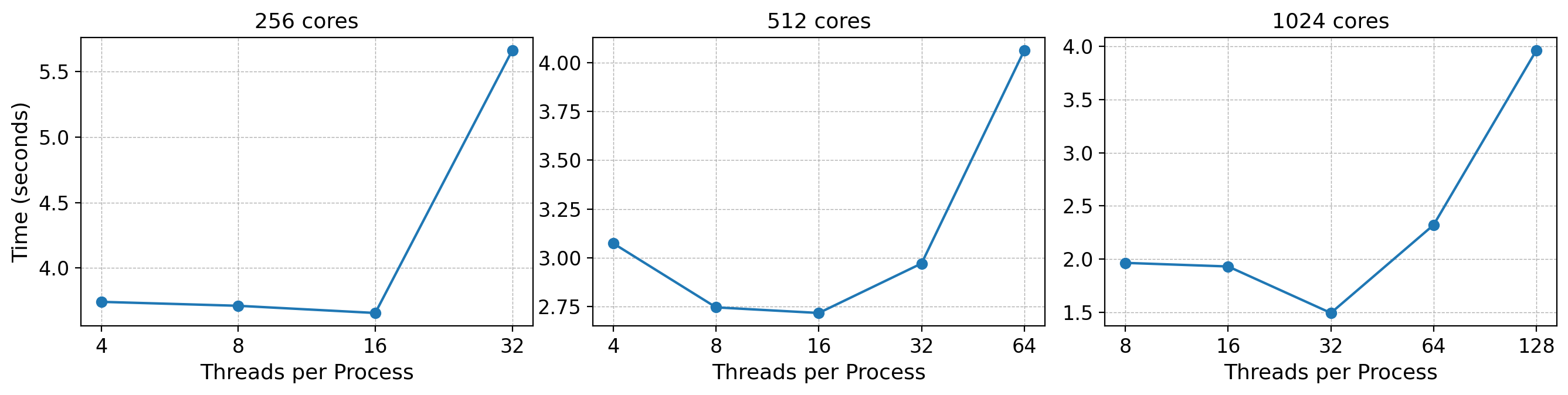}
\caption{Execution time (median over 5 runs) versus number of threads per process, given a fixed number of cores (256, 512, and 1024 cores) against Class D.}
\label{fig:process_width}
\end{figure*}
}

\newcommand{\FigLoadBalanceMPI}{
\begin{figure}[h]
\centering
\includegraphics[width=\columnwidth]{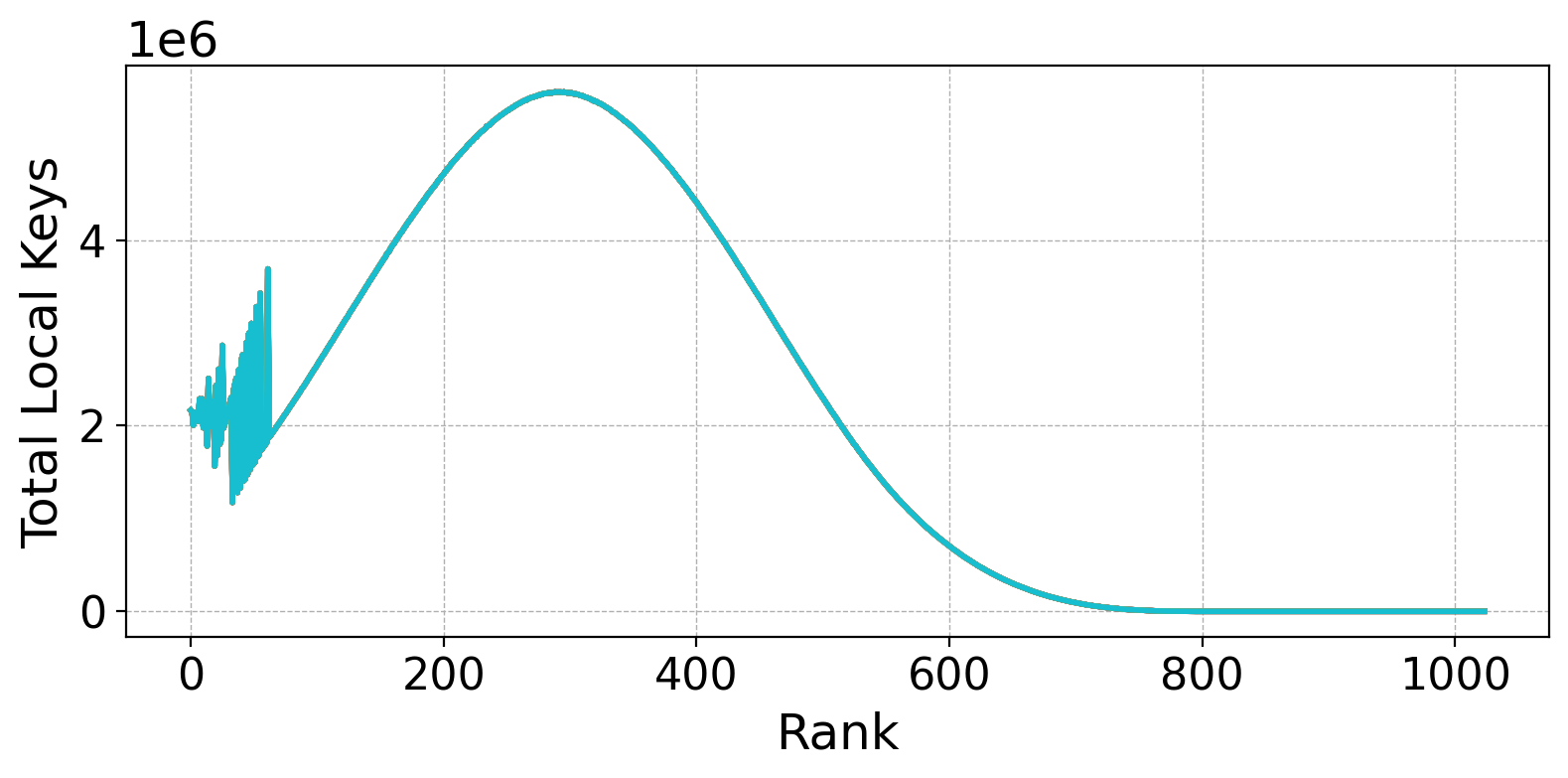}
\caption{Distribution of total local keys across 1024 processes using MPI implementation under Class D ($2^{31}$ keys, 1024 buckets). Total local keys is the number of keys held by each core after key redistribution. }
\label{fig:load_balance_mpi}
\end{figure}
}

\newcommand{\FigLoadBalanceTotalKeys}{
\begin{figure*}[t]
\centering
\includegraphics[width=\textwidth]{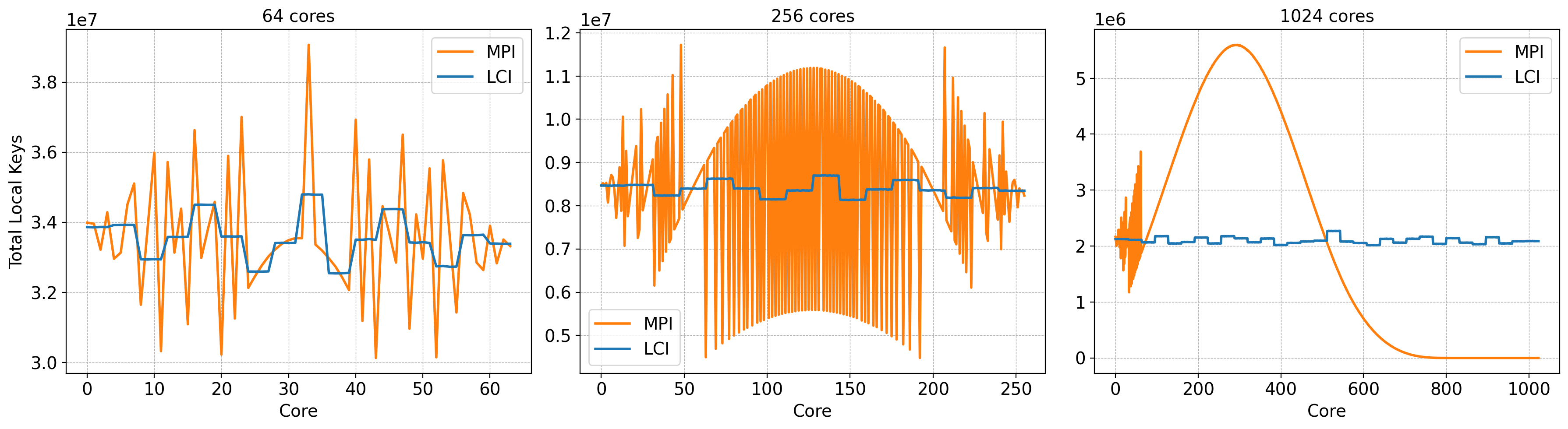}
\caption{Total local keys distribution comparison of the LCI (blue) and MPI (orange) implementations, against Class D, on 64, 256, and 1024 cores. Total local keys is the number of keys received by each core during the key exchange phase.}
\label{fig:load_balance_total_keys}
\end{figure*}
}

\newcommand{\FigLoadBalanceRcomp}{
\begin{figure}[h]
\centering
\includegraphics[width=\columnwidth]{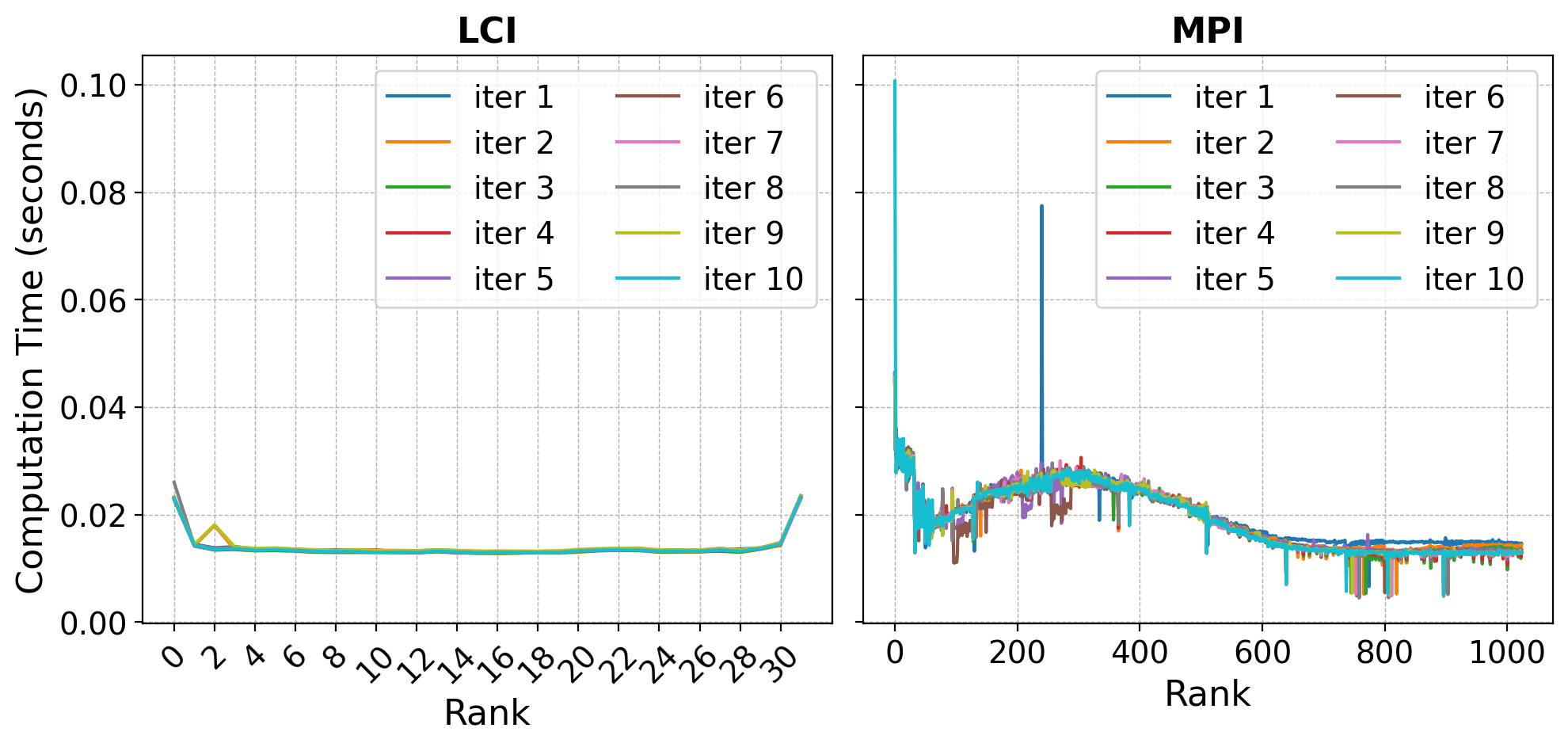}
\caption{Comparison of the computation time spent across processes under Class D, over 10 iterations of sorting, using the LCI (left) and MPI (right) implementations. Computation time is the total time spent in all computational steps (Steps 2, 3, 5, 8 in Algorithm~\ref{alg:mpi_sort}; Steps 2, 4, 6 in Algorithm~\ref{alg:lci_sort}).}
\label{fig:load_balance_rcomp}
\end{figure}
}

\newcommand{\FigLCIVariants}{
  \begin{figure}[h]
  \centering
  \includegraphics[width=\columnwidth]{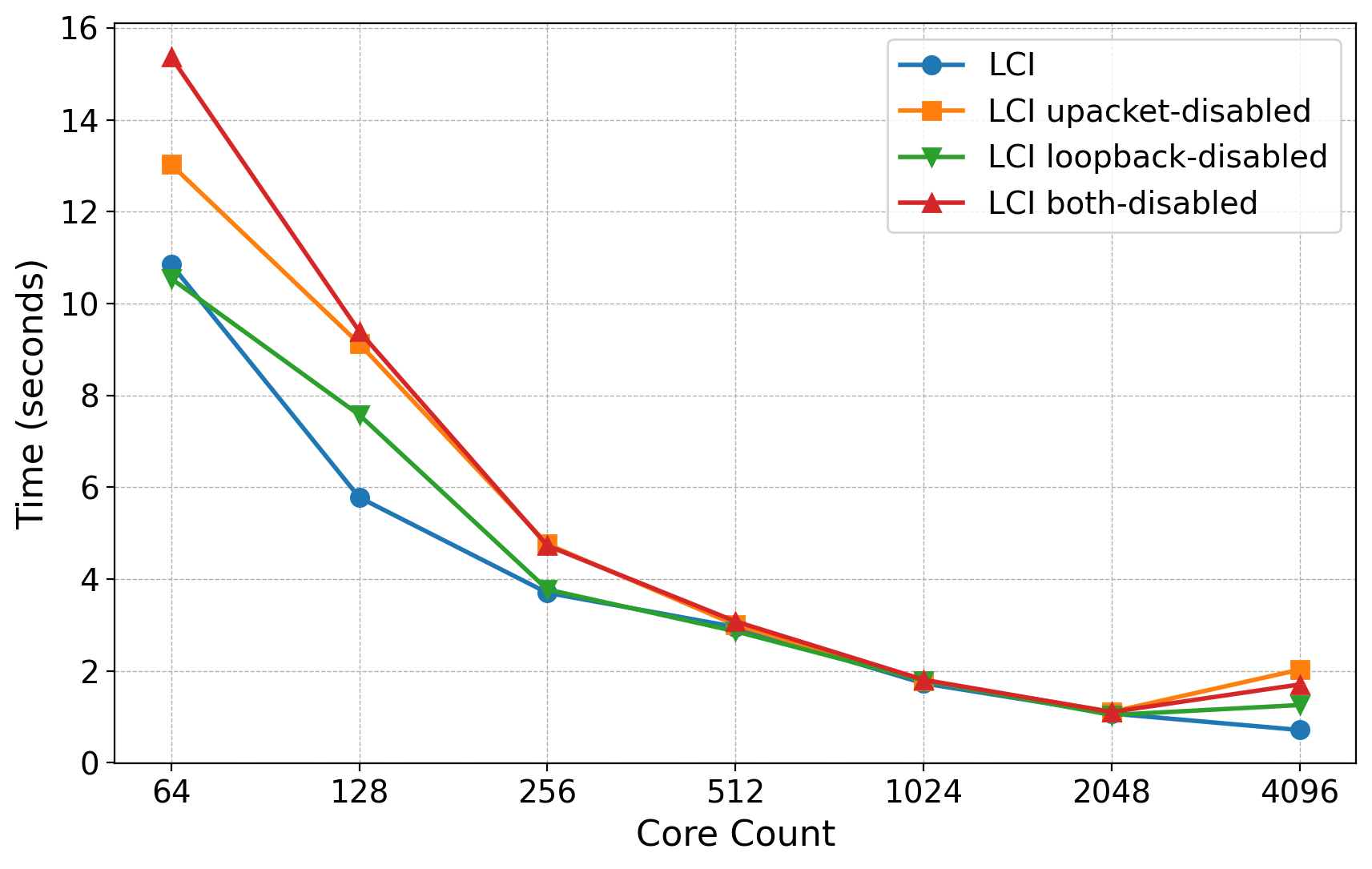}
  \caption{Execution time (median over 5 runs) of LCI variants against Class D. The variants are: (1) with loopback optimization disabled, (2) with zero-copy packet operations disabled, (3) with both enabled, (4) with both disabled.}
  \label{fig:lci_variants}
  \end{figure}
}
\section{Introduction}
\label{sec:introduction}

For decades, the bulk synchronous parallel (BSP) model has been the dominant paradigm in parallel computing, offering a predictable execution model based on lockstep phases of alternating computation and communication. However, as high-performance computing (HPC) applications scale and algorithms become increasingly irregular, the rigidity of BSP becomes a bottleneck. The requirement for global synchronization at barriers results in significant idle time during load imbalances, while its rigid structure prevents effective overlap between computation and communication.

To mitigate these issues, models that relax global synchronization have emerged. Task-based systems~\cite{kale1993CHARMPortableConcurrent,Kaiser2020HPX,bauer2012LegionExpressingLocality} allow runtimes to manage asynchrony dynamically but often impose heavy runtime overheads and require extensive refactoring of legacy code. Alternatively, the fine-grained asynchronous BSP (FA-BSP) model~\cite{sri2023fabsp} offers a middle ground. By retaining the BSP structure while enabling asynchronous communication and finer-grained actor execution, FA-BSP improves computation-communication overlap with minimal code changes.

However, current FA-BSP implementations generally rely on a single-threaded, one-process-per-core model. This approach fails to exploit modern multicore architectures effectively, as the resulting proliferation of processes increases pressure on memory subsystems and network interfaces. Furthermore, single-threaded execution cannot leverage shared-memory multithreading to mitigate load imbalances in irregular applications. This reliance on one-process-per-core is largely a historical compromise due to the poor multithreaded performance of traditional communication libraries~\cite{Patinyasakdikul2019openmpi_mt,zambre2022LessonsLearnedMPI,yan2025hpx_mpich} such as the message passing interface (MPI) \cite{mpi10} and OpenSHMEM \cite{chapman2010openshmem}.

With the advent of libraries designed for efficient multithreaded communication, such as the lightweight communication interface (LCI) \cite{yan2025lci}, these limitations can now be overcome. This paper explores the potential of multithreaded FA-BSP by presenting a case study of the integer sort (IS) benchmark from the NAS parallel benchmarks (NPB) \cite{benchmarks2006npb}. As a building block for particle simulations and graph analytics, IS is an ideal candidate for this evaluation, given its irregular communication patterns and data-dependent load imbalances arising from Gaussian key distributions.

We present a multithreaded FA-BSP implementation of Integer Sort using LCI and OpenMP, and compare it with a traditional single-threaded MPI-based BSP implementation. Our evaluation shows that combining multithreading with the FA-BSP model yields significant performance gains. This paper makes the following contributions:

\begin{enumerate} 
    \item An analysis of the inefficiencies inherent in the traditional MPI-based BSP implementation of Integer Sort.
    \item The design and implementation of a multithreaded FA-BSP algorithm of Integer Sort utilizing LCI and OpenMP.
    \item A detailed performance evaluation on a large-scale HPC system, demonstrating the specific benefits of multithreading and FA-BSP for irregular workloads. 
    \item A set of controlled experiments to identify the source of performance improvement in the LCI implementation.
\end{enumerate}

The rest of the paper is organized as follows. Section \ref{sec:related_work} provides background on the HPC programming models and communication libraries relevant to this work and prior work on Integer Sort. Section \ref{sec:background} introduces LCI and describes its features relevant to this work. Section \ref{sec:design} describes the design and implementation of our multithreaded FA-BSP Integer Sort using LCI. Section \ref{sec:evaluation} presents our performance evaluation and analysis. Finally, Section \ref{sec:conclusion} concludes the paper and discusses future work.

\section{Related Work}
\label{sec:related_work}

\subsection{Parallel Models and Communication Runtimes}

While the BSP model provides structural simplicity, decades of research have focused on addressing its limitations in handling irregular workloads. The partitioned global address space (PGAS) model \cite{el2006upc,chapman2010openshmem,bachan2019upc++} extends BSP by introducing a global shared memory abstraction and one-sided communication primitives, yet it typically retains the bulk-synchronous execution structure with global barriers. In contrast, the bulk asynchronous parallel (BASP) model \cite{Dathathri2019gluon_basp} explicitly relaxes these synchronization requirements, allowing processes to proceed immediately once local dependencies are met. More radical departures, such as task-based systems~\cite{kale1993CHARMPortableConcurrent,Kaiser2020HPX,bauer2012LegionExpressingLocality}, abandon the superstep structure entirely in favor of dynamic, graph-based execution. FA-BSP \cite{sri2023fabsp} occupies a middle ground by injecting fine-grained actor execution into the standard BSP superstep, aiming to preserve the model's simplicity while enabling the overlap of communication and computation found in asynchronous models.

However, the efficacy of these models is heavily constrained by the underlying communication libraries. Traditional message-passing libraries like MPI \cite{mpi10} and GASNet-EX \cite{bonachea_gasnet-ex_2018} were primarily optimized for single-threaded, process-based parallelism. Consequently, they often incur high lock contention when accessed by multiple threads concurrently. This bottleneck has historically forced FA-BSP implementations to rely on a one-process-per-core configuration, limiting shared-memory load balancing. Our work leverages LCI \cite{yan2025lci}, which distinguishes itself from MPI and GASNet-EX by optimizing specifically for multithreaded, fine-grained messaging, thereby enabling the multithreaded FA-BSP approach proposed in this paper. LCI has been used to improve the performance of established task-based systems~\cite{yan2025hpx_lci,mor2023PaRSEC_LCI}.

\subsection{Prior Work on Integer Sort}

The NPB IS benchmark \cite{benchmarks2006npb} has been widely studied to evaluate programming models and communication substrates. The standard NPB suite provides reference implementations in MPI (distributed memory) and OpenMP (shared memory). Several studies have attempted to optimize IS using alternative models. For example, \cite{pophale2012openshmem_is} evaluated an OpenSHMEM implementation of IS and found it resulting in slightly lower performance compared to the bulk-transfer MPI baseline.

Other works have modified the benchmark itself to avoid the load imbalancing issue. The ISx benchmark \cite{hanebutte2015isx} provides implementations in MPI, OpenSHMEM, and Chapel. While \cite{hemstad2016isx} demonstrated that OpenSHMEM could outperform MPI on ISx, the benchmark alters the key generation to a uniform distribution. This modification significantly reduces the load imbalance and communication irregularity inherent in the original Gaussian distribution of NPB IS. Similarly, \cite{Elmougy2025actor_isx} applied the FA-BSP model to ISx, showing gains over standard MPI and OpenSHMEM. However, because ISx effectively regularizes the workload, these results do not fully capture the challenges of highly irregular applications.

In contrast, this work targets the original NPB IS benchmark with its Gaussian key distribution. We directly tackle the pathological load imbalances and irregular communication patterns that multithreaded FA-BSP is theoretically best suited to solve.

\section{LCI Overview}
\label{sec:background}

Many HPC algorithms are traditionally implemented on top of MPI, which was originally designed around bulk-synchronous execution and relatively coarse-grained communication. While MPI provides strong single-thread performance and portability, its communication model and implementation choices make it less well-suited for algorithms that rely on fine-grained, highly asynchronous communication invoked concurrently by many threads. In such settings, communication progress, synchronization overheads, and contention on shared communication resources can become significant performance bottlenecks.

\begin{figure}[t]
    \centering
    \includegraphics[width=\linewidth]{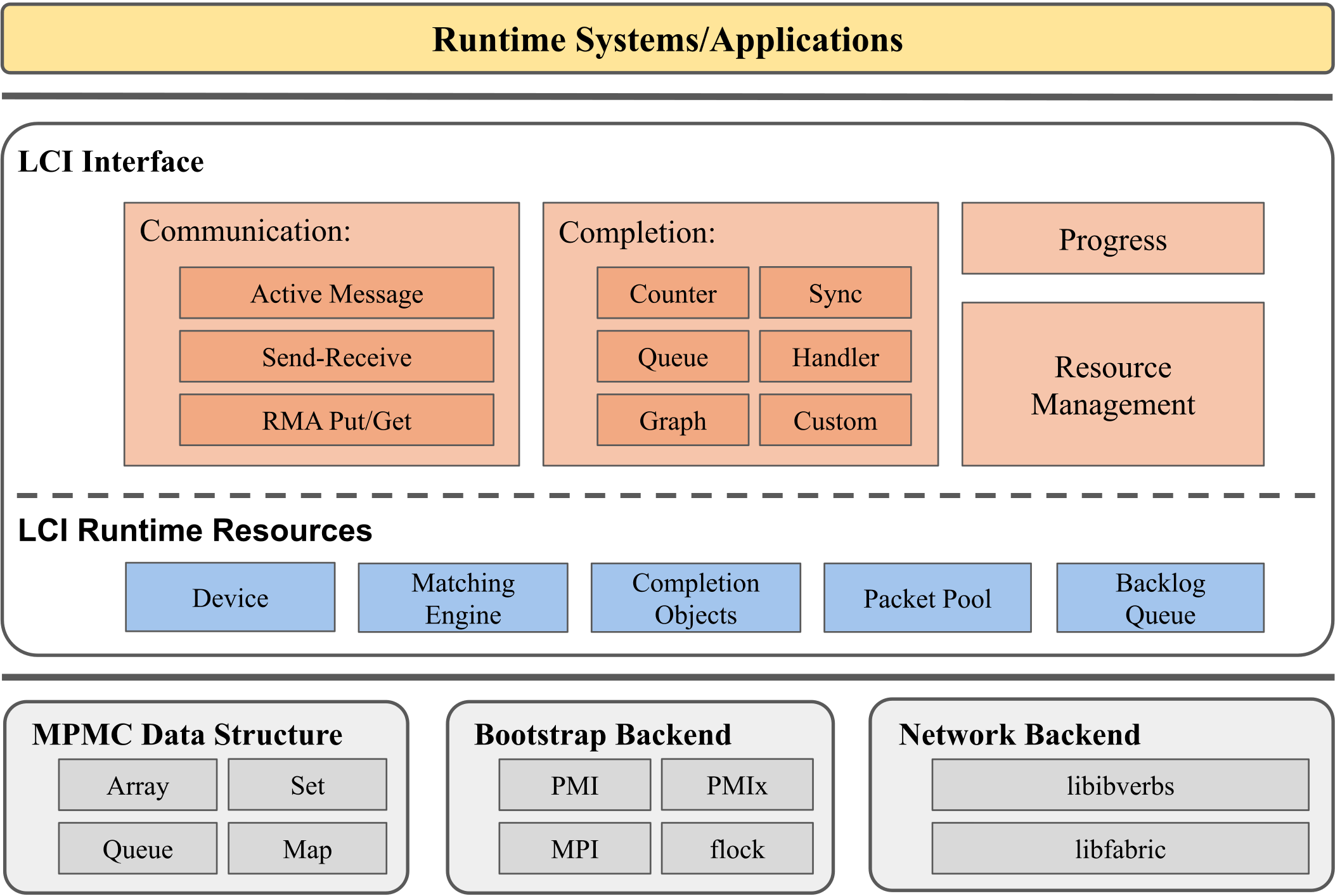}
    \caption{Overview of LCI's architecture.}
    \label{fig:lci_overview}
\end{figure}

LCI (lightweight communication interface) is a communication library designed to address these challenges by explicitly targeting asynchronous and multithreaded communication patterns. Figure~\ref{fig:lci_overview} illustrates the key components of LCI’s architecture. Rather than focusing on a specific communication paradigm, LCI supports a variety of two-sided and one-sided communication primitives along with multiple completion mechanisms that can be flexibly combined. It also provides fine-grained control over communication progress and resource mapping. These capabilities enable algorithms to aggressively overlap communication and computation, avoid global synchronization, and issue communication operations from multiple threads with different levels of resource sharing to balance thread contention and static load partitioning. As a result, LCI is particularly well-suited for algorithms that generate large numbers of small messages and rely on dynamic, irregular communication patterns. More details on LCI’s design and implementation can be found in~\cite{yan2025lci}. In the rest of this section, we briefly highlight the LCI features most relevant to our integer sorting implementation.

LCI provides several communication primitives, including tag-matching send/receive, remote memory access (RMA) put/get, active messages, and collective communication. In this work, we primarily rely on LCI’s active message primitives to implement fine-grained asynchronous communication. Active messages allow computation to be triggered directly upon message arrival, making them well suited for irregular and event-driven algorithms. Internally, LCI supports both eager and rendezvous protocols and selects between them based on message size and configuration. In our implementation, we configure LCI such that all messages use the eager protocol.

In addition to communication primitives, LCI offers multiple completion checking mechanisms, including counters, completion queues, handlers, synchronizers (similar to MPI requests), and dependency graphs. These mechanisms can be selected independently for each communication operation, allowing users to decouple communication initiation from completion handling and to integrate communication naturally into different execution models. Applications may freely combine primitives and completion mechanisms to express their desired synchronization semantics with minimal overhead. In this work, we primarily use completion counters for local completion and function handlers for remote completion.

LCI also allows users to explicitly control the mapping between communication resources and operations, enabling flexible tuning of resource replication and sharing. One key LCI resource that users frequently manage explicitly is the \emph{device}. An LCI device represents a set of communication resources (such as network queues and progress state) that can be progressed independently of other devices by distinct threads. A process may create multiple devices and assign different threads or communication operations to different devices to reduce contention. Communication progress is driven through explicit progress calls on a per-device basis from user threads.

Finally, LCI supports zero-copy eager messaging through the use of pre-registered internal buffers, referred to as packets. Applications can directly assemble send buffers or receive message payloads in these packets, eliminating intermediate memory copies that are typically required by eager messaging protocols. This capability is particularly beneficial for fine-grained communication patterns, where per-message overhead can otherwise dominate execution time.
\section{Design and Implementation}
\label{sec:design}

In this section, we describe our implementation of the Integer Sort benchmark from the NAS Parallel Benchmarks suite using fine-grained asynchronous messaging and multithreading. First, we analyze the original MPI implementation and its inherent limitations. Then, we present a detailed description of our multithreaded FA-BSP design using LCI.

\subsection{The Original MPI Implementation}
\label{sec:mpi_implementation}

Algorithm~\ref{alg:mpi_sort} outlines the key steps of the MPI-based IS implementation \cite{nasa2026npb_code}. This algorithm performs a load-balanced parallel bucket sort. 

One key factor stressed by the IS benchmark is the ability to handle irregular workloads. Accordingly, the implementation generates keys following a Gaussian distribution (Step 1), with each process holding $K_{local}$ keys. Each process then counts the number of keys in each bucket, producing $C_{local}$ (Step 2), and organizes the keys into buckets (Step 3). The global bucket sizes, $C_{global}$, are subsequently computed using \texttt{MPI\_Allreduce} (Step 4). Bucketing is performed by evenly partitioning the key space into 
$B$ equal-sized contiguous intervals; in the IS implementation, this is determined by the most significant bits of the keys.

To address load imbalance, the system dynamically assigns buckets to processes with a greedy algorithm (Step 5) to ensure a relatively even workload distribution. However, under a Gaussian distribution, the workload assigned to each process can still be unbalanced: a single bucket near the center can contain far more keys than several buckets in the tails combined. Thus, while the bucket-based load balancing is effective, it does not completely eliminate load imbalance across processes.

Once the bucket-to-process map is established, processes exchange data volume information using \texttt{MPI\_Alltoall} (Step 6) and redistribute the actual keys via \texttt{MPI\_Alltoallv} (Step 7). After redistribution, each process owns an interval of the key space and holds its portion in $K_{final}$. Step 7 becomes the bottleneck of the algorithm as the system scales. Based on our profiling using 1024 cores under Class D of the IS benchmark, Step 7 accounts for an average of 87\% of the total execution time. 

Finally, each process sorts its $K_{final}$ locally via counting sort (Step 8): a single traversal builds a histogram of key frequencies, and a prefix sum over these counts assigns each key its final global rank. 

\begin{algorithm}
\caption{Parallel Sort with MPI}
\label{alg:mpi_sort}
\begin{algorithmic}[1] % The [1] ensures every line is numbered
    \Procedure{MPI\_Parallel\_Sort}{}
        \State \textbf{Input:} $P$ (Total Processes)

        \State \textbf{Constants:} $B$ (Number of Buckets)
        
        \Statex \Comment{Generate keys - Step 1}
        \State $K_{local} \gets \text{GenerateGaussianKeys}()$
        
        \Statex
        \Statex \Comment{Count keys for each bucket - Step 2}
        \State $C_{local} \gets \text{CountBuckets}(K_{local})$
        
        \Statex
        \Statex \Comment{Organize keys into buckets - Step 3}
        \State $K_{bucketed} \gets \text{LocalBucketSort}(K_{local})$
        
        \Statex
        \Statex \Comment{Aggregate global bucket sizes - Step 4}
        \State $C_{global} \gets \texttt{MPI\_Allreduce}(C_{local}, \texttt{MPI\_SUM})$
        
        \Statex
        \Statex \Comment{Map buckets to processes (load balancing) - Step 5}
        \State $acc \gets 0$
        \State $rank \gets 0$
        \State $target\_count \gets \text{Sum}(C_{global}) / P$
        \State $S_{counts}, S_{displs}$ $\gets$ arrays of size $P$ initialized to 0
        \For{$b \in \{0, 1, \ldots, B-1\}$}
            \State $acc \gets acc + C_{global}[b]$
            \If{$acc \geq (rank + 1) \times target\_count$}
                \State $S_{counts}[rank] \gets$ CalculateSendCount()
                \State $S_{displs}[rank] \gets$ CalculateSendDispl()
                \State $rank \gets rank + 1$
            \EndIf
        \EndFor
        
        \Statex \Comment{Exchange metadata - Step 6}
        \State $R_{counts} \gets \texttt{MPI\_Alltoall}(S_{counts})$
        
        \Statex
        \Statex \Comment{Redistribute keys - Step 7}
        \State $K_{final} \gets \texttt{MPI\_Alltoallv}(K_{bucketed}, S_{counts},$ \\
        \hspace{4.9cm} $S_{displs}, R_{counts})$
        
        \Statex
        \Statex \Comment{Final Ranking via Counting Sort - Step 8}
        \State $histogram \gets \text{CountKeyFrequencies}(K_{final})$
        \State $GlobalRank \gets \text{PrefixSum}(histogram)$
    \EndProcedure
\end{algorithmic}
\end{algorithm}

This algorithm suffers from several critical limitations:

\textbf{Load Imbalance:} The single-threaded nature of the implementation forces a one-process-per-core execution model. Under this model, each process has only one core to handle its bucket(s). As a result, a core with heavy bucket(s) cannot split that work with other cores, so load imbalance persists despite effective bucket-level balancing. This imbalance is exacerbated by the BSP model: collectives like \texttt{MPI\_Alltoallv} require global participation, forcing processes with lighter workloads to idle at barriers while waiting for heavily loaded processes to complete, resulting in substantial resource wastage. 

\FigLoadBalanceMPI

Figure~\ref{fig:load_balance_mpi} visualizes the load imbalance problem that MPI implementation faces. This is sampled from a run with 1024 cores under Class D, where the benchmark fixes the number of buckets to 1024. The curve shows the distribution of the keys that each process receives after key redistribution (thus after bucket-based load balancing). While the bucket assignment balancing works in the left end of the bell curve, the heavy buckets in the middle are unable to be balanced. Under the one-process-per-core model, each of those heavy buckets is assigned to exactly one core and cannot be split across others. This leaves the major workload unbalanced. Furthermore, the last 25\% ranks receive a negligible number of keys compared to other ranks, wasting a lot of allocated computational resources. 

\textbf{Lack of Overlap:} The reliance on \texttt{MPI\_Alltoallv} for key redistribution acts as a hard barrier: processes cannot process incoming data until the entire exchange is complete. This prevents the overlap of computation and communication. Ideally, a process should begin processing a subset of keys as soon as they arrive, rather than waiting for all keys to be received.

\textbf{Data Marshalling Overhead:} \texttt{MPI\_Alltoallv} requires packing data into contiguous buffers for transmission. This necessitates at least one user-space copy (Step 3) to arrange keys into buckets. Internally, because the network interface card (NIC) can only access registered memory regions, MPI often needs to perform additional memory copies or handshakes before transferring data, further increasing latency and CPU overhead.

\textbf{Intra-node Communication Inefficiency:} In a one-process-per-core model, communication between cores on the same node must traverse the MPI stack. While MPI implementations optimize for shared memory, they still incur higher latency and overhead compared to the direct memory access available to threads within a single process.

\textbf{Memory Overhead:} \texttt{MPI\_Alltoallv} requires processes to allocate send and receive buffers large enough to accommodate the entire communication volume. In the NPB IS implementation, this results in double buffering (one buffer for locally sorted keys and another for received keys), which doubles the memory footprint during redistribution. Additionally, the one-process-per-core model increases memory pressure by replicating the MPI runtime state for every core.

\textbf{Scalability Constraints:} The algorithm's scalability is limited by the fixed number of buckets (1024 for Class D). Since the total process count cannot exceed the number of buckets, the algorithm cannot scale beyond 1024 processes (typically just 8 nodes on modern HPC systems). Furthermore, the proliferation of processes increases the cost of global collectives and bootstrapping.

\subsection{Multithreaded FA-BSP with LCI}

Our implementation retains the bucket-sort structure but replaces bulk-synchronous key exchange (\texttt{MPI\_Alltoallv}) with fine-grained asynchronous communication driven by active messages. Plus, we employ OpenMP for thread-level parallelism. While traditional libraries often struggle to support efficient multithreaded active messages, we leverage LCI to overcome these bottlenecks.

The LCI implementation of the NPB IS benchmark is built around aggregated active messages. Algorithm~\ref{alg:am_handler} outlines our active message handler. When an active message arrives at a process, a thread calling \texttt{lci::progress} picks it up and processes it. The payload of each active message is a buffer of keys. For each key $k$, the thread increments the per-key counter $histogram[k]$. The receive count $R_{global}$ of the process is also updated. These per-process variables will be used in Algorithm~\ref{alg:lci_sort}.

\begin{algorithm}
\caption{Active Message Handler}
\label{alg:am_handler}
\begin{algorithmic}[1]
    \Procedure{AM\_HANDLER}{payload} 
        \For{$k \in payload$}
            \State \textbf{atomic} $histogram[k] \gets histogram[k] + 1$
        \EndFor
        \State \textbf{atomic} $R_{global} \gets R_{global} + payload.size / keysize$
    \EndProcedure
\end{algorithmic}
\end{algorithm}

\begin{algorithm}
\caption{LCI-based Parallel Sort with Active Messages}
\label{alg:lci_sort}
\begin{algorithmic}[1]
    \algblockdefx[PAR]{Par}{EndPar}%
        {\textbf{parallel region}}%
        {\textbf{end parallel region}}
    \algblockdefx[PARFOR]{ParFor}{EndParFor}%
        [1]{\textbf{parallel for} #1 \textbf{do}}%
        {\textbf{end parallel for}}
    
    \Procedure{LCI\_Sort\_Main}{}
        \State \textbf{Input:} $my\_rank$ (Process Rank)
        \State \textbf{Constants:} $B$ (Number of Buckets)
        
        \Statex \Comment{Key generation - Step 1}
        \State $Keys \gets \text{GenerateGaussianKeys}()$
        
        \Statex
        \Statex \Comment{Count keys for each bucket - Step 2}
        \State Initialize thread-local histograms $H_{tl}$ of size $B$
        \ParFor{$k \in Keys$}
        \State $H_{tl}[\text{bucket}(k)]\text{.increment}()$
        \EndParFor
        \State \textbf{critical section} Merge $H_{tl}$ into $H_{local}$
        
        \Statex
        \Statex \Comment{Aggregate global bucket sizes - Step 3}
        \State $H_{global} \gets \texttt{lci::reduce\_x}(H_{local}, \texttt{SUM})$
        \State $H_{global} \gets \texttt{lci::broadcast\_x}(H_{global})$
        
        \Statex
        \Statex \Comment{Map buckets to processes (load balancing) - Step 4}
        \State $Map \gets \text{MapBucketsToProcesses}(H_{global})$
        \State $R_{expected} \gets \text{CalculateExpectedRecvCount}()$
        
        \Statex
        \Statex \Comment{Redistribute keys - Step 5}
        \State \textbf{atomic} $R_{global} \gets 0$
        \State \textbf{atomic} $histogram$ $\gets$ initialized key frequency array
        \Par
            \State $Buf_{tl}$ $\gets$ thread-local aggregation buffers
            \For{$k \in Keys$}
                \State $dest \gets Map[\text{bucket}(k)]$
                \State $Buf_{tl}[dest].\text{push}(k)$
                \If{$Buf_{tl}[dest].\text{is\_full}()$}
                    \If{$dest$ = $my\_rank$}
                        \State Trigger \texttt{AM\_HANDLER}($Buf_{tl}[dest]$)
                    \Else
                        \State $\texttt{lci::send\_am}(dest, $\\
                        \hspace{4cm} $\texttt{AM\_HANDLER},$\\
                        \hspace{4cm} $Buf_{tl}[dest])$
                    \EndIf
                \EndIf
            \EndFor

            \State FlushRemainingBuffers($Buf_{tl}$)
            \State \textbf{while} $R_{global} < R_{expected}$ \textbf{do}
            \State \quad $\texttt{lci::progress\_x}()$
            \State \textbf{end while}
        \EndPar

        \Statex
        \Statex \Comment{Final Ranking (Parallel Prefix Sum) - Step 6}
        \Par
            \State  $sums_{local} \gets \text{LocalScan}(histogram)$
            \State \textbf{barrier}
            \State \textbf{single thread:} 
            \State \quad $offsets \gets \text{ExclusivePrefixSum}(sums_{local})$
            \State $GlobalRank \gets \text{LocalScan}(histogram, offsets)$
        \EndPar
    \EndProcedure
    
\end{algorithmic}
\end{algorithm}

Algorithm~\ref{alg:lci_sort} outlines our LCI-based IS implementation. Initially, each process generates Gaussian-distributed keys (Step 1) exactly as in the MPI implementation. For step 2, we parallelize the original key counting phase. Each OpenMP thread counts a subset of keys into a thread-local histogram $H_{tl}$, followed by a critical section to merge them into a complete histogram $H_{local}$ for the process.

Next, the system aggregates global bucket sizes using \texttt{lci::reduce\_x} followed by \texttt{lci::broadcast\_x} (Step 3). LCI does not currently provide a dedicated allreduce primitive; this reduce-then-broadcast sequence aims to achieve the same effect as \texttt{MPI\_Allreduce} in the MPI implementation. 

At step 4, each process maps buckets to processes based on the global bucket sizes $H_{global}$. This is similar to step 5 in Algorithm~\ref{alg:mpi_sort}. Thus, the same load-balancing strategy is applied. The key difference is that we record a mapping from bucket indices to process ranks, rather than maintaining the arrays $S_{counts}$ and $S_{displs}$. During this mapping process, each process can also directly determine its expected receive count, $R_{expected}$.

Once the bucket-to-process mapping is established, the system begins the redistribution of keys (Step 5). Within each process, each OpenMP thread owns a temporary aggregation buffer $Buf_{tl}[dest]$ for each destination rank $dest$. The threads go through the input array in parallel. Each thread will push keys into their corresponding aggregation buffers (lines 19-20) and send an active message when a buffer is full (lines 21-29). We apply an optimization in which, if the destination is the local process, the thread directly invokes the active message handler shown in Algorithm~\ref{alg:am_handler}. After going through all the keys, the threads will flush their remaining buffers (line 31) and start handling the received active messages (lines 32-34). Until the expected receive count $R_{expected}$ is reached, each thread calls \texttt{lci::progress\_x} to progress communication, picking up active messages received by the process. Therefore, in step 5, the communication-related computation interleaves with the communication within each thread, and different threads can be working at different stages at the same time. This design can efficiently overlap communication and computation for each process.

During key redistribution, each process has already counted the key frequencies into $histogram$. To determine the global rank of each key, each process must still compute the prefix sum of these frequencies (Step 6). We implement the prefix sum in parallel. Each OpenMP thread first sums the frequencies over its statically scheduled chunk of the key range and stores the result in a shared array $sums_{local}$. After all threads have finished, a single thread computes the exclusive prefix sum of these per-thread sums in place, so each entry becomes the starting offset for the corresponding thread. Each thread then performs a sequential scan over its chunk of $histogram$, adding each frequency to its offset and writing the running total into the global rank array. The result is the inclusive prefix sum of the frequencies, i.e., the final rank of each key value.

In summary, our multithreaded fine-grained algorithm employs a range of techniques to address the limitations faced by the MPI implementation:

\textbf{Multithreading:} We use OpenMP to parallelize all major computational steps, as well as the most time-consuming key redistribution step. Running multiple threads per process addresses \emph{load imbalance} by letting many cores share a process's buckets. Workloads inside a bucket can be distributed evenly across threads (e.g., during counting, sending, and receiving loops) so that heavy buckets no longer stall a single core while others idle. It also reduces \emph{memory overhead} by requiring fewer processes for a fixed number of cores, cutting replication of runtime state and reducing the number of send/receive buffers. In addition, it relaxes \emph{scalability constraints} by allowing more cores to be used without increasing the process count. The 1024-bucket limit constrains the number of processes rather than the total number of cores, and smaller process counts also help keep collective and bootstrap costs low.

\textbf{Fine-grained Asynchronous Communication:} We replace the synchronous \texttt{MPI\_Alltoallv}, along with its requisite buffer preparation and subsequent unpacking, with a continuous flow of asynchronous active messages. In the redistribution phase, threads iterate through the input array and send keys to their destination immediately (lines 18-30 in Algorithm \ref{alg:lci_sort}). Upon arrival, an active message handler atomically updates the local counters (Algorithm \ref{alg:am_handler}). This design eliminates the global synchronization barrier imposed by \texttt{MPI\_Alltoallv} and thereby addresses the \emph{lack of overlap} between communication and computation. Threads can continue sending keys while concurrently handling incoming messages, allowing communication progress and counter updates to proceed in parallel. The approach also reduces \emph{data marshalling overhead}, since it avoids the bulk packing and unpacking required by \texttt{MPI\_Alltoallv}. Instead, keys are streamed in small messages, and the active message handler consumes them immediately without using a separate receive buffer. In addition, the design lowers \emph{memory overhead} by eliminating full-sized send and receive buffers at each process. We instead use smaller per-thread, per-destination buffers (LCI packets) that are recycled inside LCI as messages are sent, while incoming data is processed directly in the handler rather than staged in a large receive buffer. 

\textbf{Thread-local Aggregation:} Fine-grained sending could incur excessive per-message overhead if each key were sent alone. To keep the cost low, each thread uses per-destination aggregation buffers ($Buf_{tl}$): keys are accumulated until a buffer (64 KB by default) is full, after which the full buffer is sent. This size balances memory use and bandwidth efficiency.

\textbf{Zero-copy Eager Messages:} For high-performance interconnects (e.g., InfiniBand, Slingshot-11), communication buffers must be registered with the NICs. LCI manages a pool of pre-registered "packets" for this purpose. Typically, libraries copy user data into these packets (eager protocol). To save the overhead of this copy, we utilize LCI's \texttt{lci::get\_upacket} API to acquire a pointer to an internal packet, allowing threads to assemble the messages directly into the NIC-registered memory buffers. This enables true zero-copy sending. On the receiver side, LCI delivers the internal packet directly to the handler, avoiding a receive-side copy as well. This addresses \emph{data marshaling overhead} by eliminating the send- and receive-side copies that would otherwise occur with user-provided buffers.

\textbf{Loopback Optimization:} Keys destined for the local process bypass the network stack entirely (Algorithm~\ref{alg:lci_sort}, lines 22-23). For these keys, the active message handler is invoked directly via function call, leveraging shared memory to minimize latency. Thus, this addresses the \emph{intra-node communication inefficiency}.

Table \ref{tab:comparison} summarizes how each of the specific limitations of the MPI implementation can be addressed by our design choices.

\begin{table}[ht]
\centering
\small
\caption{Comparison of Limitations and Solutions in LCI-based Implementation}
\label{tab:comparison}
\begin{tabular}{p{3.5cm}p{4.5cm}}
\hline
\textbf{Limitation} & \textbf{Key Techniques} \\
\hline
Load imbalance & Multithreading \\
Missed opportunities for overlap & Fine-grained communication \\
Data marshalling overhead & Fine-grained communication, zero-copy packet API \\
Intra-node communication inefficiency & Loopback optimization \\
Memory overhead & Multithreading, Fine-grained communication \\
Scalability constraints & Multithreading \\
\hline
\end{tabular}
\end{table}

We note that our LCI-based multithreaded fine-grained algorithm introduces certain complexities that are absent in the MPI implementation and in other asynchronous models such as single-threaded FA-BSP. In our design, atomic operations are used to update counters within the active message handler, which may introduce contention among threads. This overhead represents a trade-off of the multithreaded approach. In contrast, single-threaded models such as ActorISx \cite{Elmougy2025actor_isx} avoid such contention but do not benefit from the improved load balancing enabled by multithreading. 

Additionally, the computational steps must be explicitly parallelized using OpenMP, and cache locality may be reduced because keys are no longer pre-sorted into local buckets. Furthermore, since LCI currently does not optimize intra-node inter-process communication, data transfers between processes on the same node may be less efficient than in MPI.

Section~\ref{sec:evaluation} presents a detailed performance evaluation comparing our LCI-based implementation with the original MPI version. Overall, we find that the benefits of multithreading and fine-grained asynchronous communication outweigh these complexities, resulting in significant performance improvements.

The source code of our LCI implementation is publicly available at https://github.com/Ekiiiim/IntegerSort-LCI.

\section{Performance Evaluation}
\label{sec:evaluation}

\subsection{Experimental Setup}

We evaluate our multithreaded FA-BSP implementation against the NPB3.4.3 MPI implementation \cite{nasa2026npb_code}. Experiments are conducted on the Purdue Anvil supercomputer~\cite{anvil}, which uses the InfiniBand network. Table~\ref{tab:platform_config} shows the detailed platform configurations.

\begin{table}[h]
\caption{Platform Configurations}
\label{tab:platform_config}
\centering
\small
\begin{tabular}{lc}
\hline
\textbf{Platform} & Purdue Anvil \\
\hline
\textbf{CPU Model} & AMD EPYC 7763 \\
\textbf{Sockets per Node} & 2 \\
\textbf{Cores per Socket} & 64 \\
\textbf{Total Cores per Node} & 128 \\
\textbf{Memory per Node} & 256 GB \\
\textbf{Network} & HDR InfiniBand \\
\textbf{Network Bandwidth} & 100 Gbps \\
\hline
\end{tabular}
\end{table}

The experiments use the Class D ($2^{31}$ keys, 1024 buckets) and Class E ($2^{35}$ keys, 1024 buckets) problem sizes of NPB3.4.3. We run each experiment five times and report the median time. As specified in the IS benchmark implementation \cite{nasa2026npb_code}, each run includes 10 iterations of sorting, and the maximum time across iterations is reported as the time for the run. Key generation is excluded from timing. We always run the MPI implementation with one process per core. For the LCI version, threads are pinned to cores. We use the current LCI master branch (Hashtag \texttt{f5f503}) and OpenMP 4.5. We compile the MPI implementation with system-installed Open MPI 4.0.6. The system-installed Open MPI is compiled with the UCX backend (version 1.18.0). LCI is compiled with its libibverbs backend. 

\subsection{Overall Performance}
\label{sec:overall_performance}

\FigOverall

\FigProcessWidth

Figure~\ref{fig:overall} shows the strong scaling results of the two implementations on a range of core counts from 64 to 4096 (from $0.5$ to 32 nodes). For this experiment, we allocate 1 LCI device per thread. We configure the number of processes and threads as follows. For Class D, we configure threads per process $t_D(c)$ as a function of total core count $c$:
\begin{equation}
t_D(c) = \begin{cases}
4 & \text{if } c = 64 \\
16 & \text{if } 64 < c < 1024 \\
32 & \text{if } 1024 \leq c < 4096 \\
64 & \text{if } c = 4096
\end{cases}
\end{equation} 
The number of processes is then calculated as $p_D(c) = c/t_D(c)$. For Class E, the IS benchmark requires the minimum process number to be 64, so we configure threads per process $t_E(c)$ as follows:
\begin{equation}
t_E(c) = \begin{cases}
4 & \text{if } c = 256 \\
8 & \text{if } c = 512 \\
16 & \text{if } c = 1024 \\
32 & \text{if } c > 1024
\end{cases}
\end{equation}
These configurations were found to be optimal in our process width experiments, which are detailed in Section~\ref{sec:eval_config}.

On a single node in Class D, the MPI implementation outperforms LCI. This advantage stems from MPI’s mature optimizations for intra-node shared-memory communication, as well as the additional overhead introduced by the OpenMP-based multithreading in the LCI implementation. At this scale, the disadvantages of MPI’s one-process-per-core model have not yet become significant.

However, as the core count increases, the LCI implementation demonstrates superior scaling. In Class D, the MPI implementation stops scaling beyond 512 cores. In contrast, the LCI implementation continues to scale to 4096 cores (32 nodes). At 1024 cores (8 nodes), LCI achieves a 56.9\% reduction in median execution time compared to MPI. In Class E, the LCI implementation shows the same scaling behavior as in Class D.

Notably, the MPI implementation cannot run beyond 1024 cores because the number of buckets in the NPB IS benchmark is hardcoded at 1024 and the process count cannot exceed the bucket count. This limitation restricts the MPI implementation from leveraging additional parallelism at higher core counts. In contrast, the LCI implementation's multithreaded design allows it to utilize multiple threads per process, effectively lifting the constraint of the bucket count. This reflects a common scalability limitation of the traditional one-process-per-core model.

\subsection{LCI Configuration Analysis}
\label{sec:eval_config}

We analyze the sensitivity of LCI performance to key configuration parameters.

\textbf{Process width:} 
Recall that in Section~\ref{sec:overall_performance}, we used a specific configuration of process width (number of threads per process), defined by equations (1) and (2). Our experiments on process width show that those configurations generally offer the best performance. 

Figure~\ref{fig:process_width} shows the experimental results for three representative core counts: 256, 512, and 1024 cores (2, 4, and 8 nodes). We observe that an intermediate process width offers the best performance.

This behavior reflects a fundamental trade-off between process count and per-process thread count. With too many processes and too few threads per process, the system approaches the one-process-per-core model. This would lead to the problems faced by MPI implementation, such as load imbalance and inter-process communication overhead. Conversely, with too few processes and too many threads per process, the system would suffer from thread contention and reduced memory locality. When the process is too wide, its threads need to access memory across non-uniform memory access (NUMA) nodes. 

The results indicate that the optimal configuration typically lies midway between the two extremes. In our case, the optimal point occurs when the number of threads per process is approximately equal to the number of processes.

\textbf{Device Count:} The LCI implementation can utilize a configurable number of LCI devices per process to further boost the multithreading communication efficiency. We vary the number of devices per process to study its impact on the performance. Figure~\ref{fig:device_count} shows the results. 

\FigDeviceCount

\FigLoadBalanceTotalKeys

We observe no significant differences in execution time across the different LCI device configurations. This can be attributed to the efficient intra-device multithreading parallelism in the LCI runtime, which is implemented with finer-grained locking and more extensive use of atomic operations than typical MPI implementations.
%This can be interpreted in two ways: (1) \emph{thread contention on LCI devices is small}, i.e., multiple threads can share a device without noticeably increasing serialization or synchronization overhead; and (2) \emph{the underlying communication stack is efficient}, i.e., InfiniBand and LCI devices provide sufficient per-device throughput and low latency so that adding more devices does not further reduce time in this setting (the bottleneck may lie elsewhere).

\subsection{Load Balancing Analysis}
\label{sec:load_balancing}

The irregular nature of the Gaussian key distribution in NPB IS creates inherent load imbalances. In Section~\ref{sec:mpi_implementation}, we briefly discussed how the MPI implementation suffers from the unbalanced keys per core. Here, we analyze how the multithreaded FA-BSP implementation mitigates this load balancing problem.

Figure~\ref{fig:load_balance_total_keys} illustrates the load-balancing effect of the LCI implementation. We measure the total number of keys received by each core in both implementations. We collect results for Class D on a range of core counts from 64 to 1024 and show the representative core counts in the figure. We observe that multithreading effectively redistributes workload across cores, and this effect is more prominent at higher core counts.

The improved load balancing arises from a fundamental design difference. The LCI implementation allows multiple cores to work on each bucket, whereas the MPI implementation cannot. With one process per core, each bucket in the MPI implementation is handled by a single core; when a bucket is heavily loaded, that core becomes overloaded.

Consider a configuration with 1024 cores. Under the one-process-per-core model, this requires 1024 processes, resulting in 1 core being assigned to each bucket. With a Gaussian-like key distribution, heavily populated buckets overload their assigned cores, while lightly populated buckets leave many cores underutilized.

In contrast, the multithreaded implementation can use 32 processes with 32 threads per process. Each process is responsible for multiple buckets, and multiple cores can share the work within each bucket. This ability to parallelize computation within buckets gives our implementation greater flexibility and enables more effective load balancing across cores.

The benefits of multithreading are also evident in the distribution of computation time across processes, which is significantly more balanced in the multithreaded configuration.

\FigLoadBalanceRcomp

Figure~\ref{fig:load_balance_rcomp} shows the local computation time of each process for the two implementations. We observe that improved load balancing leads to a more even distribution of computation time across processes. Across all experiments, the multithreaded implementation also avoids the irregular peaks in computation time observed in the MPI implementation.

\subsection{Communication Overhead Breakdown}

To isolate the benefits of fine-grained communication and LCI-specific features, we conduct controlled experiments with different configuration options:

\textbf{Loopback Optimization}: Our loopback optimization allows keys destined for the local process to bypass the network stack entirely and invoke the active message handler directly via function calls. This design avoids the additional communication-stack overhead inherent in the one-process-per-core model, even when advanced inter-process communication optimizations are enabled. By disabling this optimization, we study the performance overhead introduced by the communication stack.

\textbf{Zero-Copy Packet Operations:} LCI’s zero-copy active message feature allows threads to assemble aggregated messages directly in LCI’s internally pre-registered buffers, thereby eliminating the memory copies typically required by eager messaging protocols. By disabling zero-copy packet operations, we study the performance impact of these additional memory copies.

\FigLCIVariants

Figure~\ref{fig:lci_variants} shows the experimental results. We observe that enabling both LCI configurations is helpful for the performance, especially at a low core count (within 2 nodes) or a high core count (at least 32 nodes). However, for core counts 512-2048, the performance is very similar across all variants, implying that the bottleneck is somewhere else. Nonetheless, the results suggest that enabling both configurations has a real impact on application performance.

\section{Conclusion}
\label{sec:conclusion}

We have presented a multithreaded FA-BSP design for Integer Sort, implemented using LCI and OpenMP.
Our results show that this design is effective for the Integer Sort benchmark and achieves better performance and scalability than the NPB 3.4.3 MPI-based BSP implementation.

This work showcases the benefits of asynchronous multithreaded FA-BSP for irregular parallel applications in general. For workloads with load imbalance, such as the Gaussian-like key distribution in Integer Sort, fine-grained asynchrony and multithreading help reduce the impact of uneven work distribution. The multithreaded design allows threads within a process to make independent progress and provides more flexibility in how work is mapped to cores.

In summary, our study shows that asynchronous multithreaded FA-BSP is a practical and effective model for irregular parallel workloads. Future work includes applying this design to other irregular algorithms and further exploring adaptive thread, process, and device configurations.

\section*{Acknowledgment}

This work used Anvil~\cite{anvil} at Rosen Center for Advanced Computing through allocation CIS250465 from the Advanced Cyberinfrastructure Coordination Ecosystem: Services \& Support (ACCESS) program, which is supported by U.S. National Science Foundation grants \#2138259, \#2138286, \#2138307, \#2137603, and \#2138296.

\bibliographystyle{IEEEtran}
\bibliography{reference}

\end{document}